# Heat Dissipation in Atomic-Scale Junctions

*Woochul Lee[1, †], Kyeongtae Kim[1, †], Wonho Jeong[1], Linda Angela Zotti[2], Fabian Pauly[3],*

*Juan Carlos Cuevas[2, *], Pramod Reddy[1, 4, *]*

[1] Department of Mechanical Engineering, University of Michigan, Ann Arbor, Michigan, 48109, USA

[2] Departamento de Física Teórica de la Materia Condensada, Universidad Autónoma de Madrid, Madrid, 28049, Spain

[3] Department of Physics, University of Konstanz, D-78457 Konstanz, Germany

[4] Department of Materials Science and Engineering, University of Michigan, Ann Arbor, Michigan, 48109, USA

[†]These authors contributed equally to this paper.

*e-mail: *pramodr@umich.edu, juancarlos.cuevas@uam.es*

**Atomic and single-molecule junctions represent the ultimate limit to the miniaturization of electrical circuits[1]. They are also ideal platforms to test quantum transport theories that are required to describe charge and energy transfer in novel functional nanodevices. Recent work has successfully probed electric and thermoelectric phenomena[2-8] in atomic-scale junctions. However, heat dissipation and transport in atomic-scale devices remain poorly characterized due to experimental challenges. Here, using custom-fabricated scanning probes with integrated nanoscale thermocouples, we show that heat dissipation in the electrodes of molecular junctions, whose transmission characteristics are strongly dependent on energy, is asymmetric, *i.e.* unequal and dependent on both the bias polarity and the identity of majority charge carriers (electrons vs. holes). In contrast, atomic junctions whose transmission characteristics show weak energy dependence do not exhibit appreciable asymmetry. Our results unambiguously relate the electronic transmission characteristics of atomic-scale junctions to their heat dissipation properties establishing a framework for understanding heat dissipation in a range of mesoscopic systems where transport is elastic. We anticipate that the techniques established here will enable the study of Peltier effects at the atomic scale, a field that has been barely explored experimentally despite interesting theoretical predictions[9-11]. Furthermore, the experimental advances described here are also expected to enable the study of heat transport in atomic and molecular junctions—an important and challenging scientific and technological goal that has remained elusive[12,13].**

Charge transport is always accompanied by heat dissipation (Joule heating). This process is well understood at the macroscale where the power dissipation (heat dissipated per unit time) is volumetric and is given by $j^2\rho$, where $j$ is the magnitude of the current density and $\rho$ is the electrical resistivity. Heating in atomic-scale junctions is expected to be fundamentally different as charge transport through such junctions is largely elastic[14,15], *i.e.* without exchange of energy in the contact region. Recent experiments have probed the local non-equilibrium electronic and phononic temperatures in molecular junctions[16-18] to obtain insights into the effect of electron-electron and electron-phonon interactions on heat dissipation at the atomic scale. However, experimental challenges in quantitatively measuring atomic-scale heat dissipation have impeded the elucidation of a fundamental question: What is the relationship between the electronic transmission characteristics of atomic and molecular junctions (AMJs) and their heat dissipation properties?

In this work, we overcome this challenging experimental hurdle by leveraging custom-fabricated Nanoscale–Thermocouple Integrated Scanning Tunneling Probes (NTISTPs) shown in Figs. 1a & b. The NTISTPs feature an outer gold (Au) electrode that is electrically isolated but thermally well connected to the integrated gold-chromium thermocouple via a thin (70 nm) silicon nitride film (see supplementary information (SI) for fabrication details). In order to probe heat dissipation we first created a series of AMJs (see Fig. 1c) between the outer Au electrode of the NTISTP and a flat Au substrate. Application of a voltage bias across such AMJs results in a temperature rise of the integrated thermocouple due to heat dissipation in the NTISTP's apex on a length scale comparable to the inelastic mean free path of electrons in Au[19]. The power dissipation in the probe ($Q_P$) and the temperature rise of the thermocouple ($\Delta T_{TC}$), located ~300 nm away from the apex, are directly related by $Q_P = \Delta T_{TC} / R_P$ (see Methods), where $R_P$ is the thermal resistance of the NTISTP (see Fig. 1b). Further, $\Delta T_{TC}$ is related to the thermoelectric voltage output of the thermocouple ($\Delta V_{TC}$) by $\Delta V_{TC} = -S_{TC} \times \Delta T_{TC}$, where $S_{TC}$ is the effective

Seebeck coefficient of the thermocouple. We note that $R_P$ and $S_{TC}$ were experimentally determined to be 72800 ± 500 K/W and 16.3 ± 0.2 µV/K, respectively (see SI).

We began our experimental studies, at room temperature, by trapping single molecules of 1,4-benezenediisonitrile (BDNC, see Fig. 1c) between the Au electrodes of the NTISTP and the substrate using a break junction technique[5,20]. We first obtained electrical conductance versus displacement traces by monitoring the electrical current under an applied bias while the NTISTP-substrate separation was systematically varied. Figure 2a shows representative conductance traces along with a histogram obtained from 500 such curves. The histogram features a peak at ~$0.002G_0$ ($G_0 = 2e^2/h$ ~$(12.9\ k\Omega)^{-1}$), which represents the most probable low-bias conductance of Au-BDNC-Au junctions ($G_{Au-BDNC-Au}$) and is in good agreement with past work[21].

In order to probe heat dissipation we created stable Au-BDNC-Au junctions with a conductance that is within 10% of the most probable low-bias conductance[20]. We studied heat dissipation in 100 distinct Au-BDNC-Au junctions, at each bias, to obtain the average temperature rise ($\Delta T_{TC,\ Avg}$) and the time-averaged power dissipation in the NTISTP ($Q_{P,\ Avg}$) for both positive and negative biases. Here, a positive (negative) bias corresponds to a scenario where the probe is grounded, while the substrate is at a higher (lower) potential. We note that a modulated voltage bias was applied to the junctions to obtain $\Delta T_{TC,\ Avg}$—with high resolution— for both positive and negative biases (see Methods and SI). This modulation scheme enables rejection of broadband noise and plays a critical role in performing high-resolution thermometry. The circles (triangles) in Fig. 2b represent the measured $\Delta T_{TC,\ Avg}$ as well as the estimated $Q_{P,\ Avg}$ for positive (negative) biases as a function of the total time-averaged power dissipation in the junctions ($Q_{Total,\ Avg} = (I \times V)_{Avg}$) at each bias voltage. Here, $V$ is the applied bias and $I$ is the resultant electric current through the junction. We note that the current-voltage (I-V) characteristics of Au-BDNC-Au junctions are non-linear (Fig. 2c), therefore, in general $Q_{Total,\ Avg} \neq G_{Au-BDNC-Au}V^2$. The dotted line corresponds to the expected temperature rise of the probe if the heating was symmetric, *i.e.* if half of the total power was dissipated in the probe ($\Delta T_{Symmetric} = Q_{Total,\ Avg}/2R_P$). It can be clearly seen that for a given $Q_{Total,\ Avg}$ the power dissipation in the probe is larger under a negative bias than a positive bias. We also conclude that the time-averaged power dissipation in the substrate, $Q_{S,\ Avg}$, is smaller under a negative bias than under a positive bias since $Q_{P,\ Avg} + Q_{S,\ Avg} = Q_{Total,\ Avg}$. In order to clarify the voltage biases used in the experiments in the inset of Fig. 2b we present $\Delta T_{TC,\ Avg}$ as a function of the magnitude of the applied voltage bias. These results unambiguously demonstrate that heat dissipation in the electrodes of Au-BDNC-Au junctions is bias polarity dependent and unequal.

This observation raises an important question: Why is the heat dissipation in the electrodes unequal in spite of the symmetric geometry of the molecular junctions? To address this question we resort to the Landauer theory of quantum transport, which has successfully described charge transport in numerous nanostructures[19]. Within this theory, the power dissipated in the probe and the substrate, $Q_P(V)$ and $Q_S(V)$, respectively, is given by[22]:

$$Q_P(V) = \frac{2}{h}\int_{-\infty}^{\infty}(\mu_P - E)\tau(E,V)[f_P - f_S]dE \quad \text{and} \quad Q_S(V) = \frac{2}{h}\int_{-\infty}^{\infty}(E - \mu_S)\tau(E,V)[f_P - f_S]dE. \quad (1)$$

Here, $\mu_P$ and $\mu_S$ are the chemical potentials of the probe and substrate electrodes, respectively, $f_{P/S}$ represent the Fermi-Dirac distribution of the probe/substrate electrodes, and $\tau(E,V)$ is the energy ($E$) and voltage bias ($V$) dependent transmission function. Equation (1) suggests that the power dissipation in the two electrodes is, in general, unequal, *i.e.* $Q_P(V) \neq Q_S(V)$, and bias polarity dependent, *i.e.* $Q_{P/S}(V) \neq Q_{P/S}(-V)$. Specifically, it is straightforward to show that:

$$Q_P(V) - Q_P(-V) \approx 2GTSV + O(V^3) \quad \text{and} \quad Q_P(V) - Q_S(V) \approx 2GTSV + O(V^3). \quad (2)$$

Here, $G$ is the low bias electrical conductance of the junctions, $T$ is the absolute temperature, and $S$ is the Seebeck coefficient of the junction, whose sign is related to the first energy derivative of the zero-bias transmission $\tau'(E = E_F, V = 0)$ at the Fermi energy ($E_F$) resulting in a positive Seebeck coefficient for a negative first derivative and vice versa[23]. In order to test if the observed heating asymmetry can be understood within this framework, we computed $\tau(E, V = 0)$ for Au-BDNC-Au junctions using a transport method[24] based on density functional theory (DFT) (see Methods). The computed transmission function (Fig. 2d) exhibits a positive slope at the Fermi energy, in agreement with past work[25], indicating a negative Seebeck coefficient, which by virtue of equation (2) leads to higher power dissipation in the NTISTP when negative voltages are applied to the substrate. Further, the solid lines in Fig. 2b represent the relationship between $Q_P$ and $Q_{Total}$ ($Q_P + Q_S = Q_{Total}$) as computed from equation (1) under the assumption that $\tau(E,V)$ is well approximated by $\tau(E, V = 0)$. Notice that although our DFT approach overestimates the linear conductance, it describes correctly the relationship between $Q_P$ and $Q_{Total}$. The reasons for this agreement are discussed further in the SI, where we show in particular that this relation is relatively insensitive to the details of the junction geometry. The good agreement of the computed and measured relation between power dissipations provides strong support to the applicability of the Landauer theory of heat dissipation at the atomic scale.

In order to conclusively prove the relationship between electronic structure and heat dissipation, we performed additional studies in 1,4-benzenediamine (BDA, see Fig. 1c) junctions, which are expected to exhibit hole-dominated electrical transport as suggested by our calculations (Fig. 3d) and past experiments[26]. Following a procedure similar to that described above we first determined that the most probable low-bias conductance of Au-BDA-Au junctions was ~$0.005G_0$ (Fig. 3a), a value consistent with past work[27]. Measurements of heat dissipation in BDA junctions (Fig. 3b) show a remarkably different asymmetry. In particular, the BDA junctions show larger power dissipation in the probe for a positive bias than for a negative one—in strong contrast to that observed in BDNC junctions. To understand this important difference we computed the transmission function of the Au-BDA-Au junction displayed in Fig. 3d, which shows that $\tau'(E=E_F, V=0)$ is negative resulting in a positive Seebeck coefficient. This, in turn, leads to larger power dissipation in the NTISTP at positive biases. Further, the computed relationship between $Q_P$ and $Q_{Total}$ is in good agreement with our experimental observations (solid lines in Fig. 3b).

Finally, to prove the fact that no appreciable asymmetries are obtained if the transmission is weakly dependent on energy, we studied heat dissipation in Au-Au atomic junctions. We began our analysis by studying the conductance of Au-Au atomic junctions which were found to have a most probable conductance of ~$G_0$, in accordance with past studies[5,28] (see SI). Subsequently, we created 100 Au-Au atomic junctions with a low-bias conductance of $G_0 \pm 0.1G_0$ and probed heating in them. The measured $\Delta T_{TC, Avg}$ (Fig. 4a) is seen to be proportional to $Q_{Total, Avg}$ and is identical for both positive and negative biases (within experimental uncertainty (~0.1 mK)) clearly demonstrating that there is no detectable asymmetry in the power dissipation. Further, additional experiments performed at larger values of $Q_{Total, Avg}$ also show no detectable asymmetry (see inset of Fig. 4a).

Symmetric heat dissipation is indeed expected in Au-Au atomic junctions due to the weak energy dependence of their transmission function[29] which is reflected in the fact that their average thermopower vanishes[6]. In Fig. 4b we present the computed zero-bias transmission, corresponding to the Au-Au atomic junction shown in inset-i. The transmission is practically energy independent over 1 eV around the Fermi energy. This weak energy dependence results in symmetric power dissipation (from Eqs. 1 and 2) as well as linear I-V characteristics as evidenced by the experimentally obtained I-V curves shown in inset-ii of Fig. 4b.

The good agreement between the measured and computed asymmetries in the heat-dissipation characteristics of AMJs unambiguously confirms that heat dissipation is indeed intimately related to the transmission characteristics of the junctions, as predicted by the Landauer theory. We note that our results contradict recent claims[30] of asymmetric heat dissipation in Au atomic junctions, which are not in agreement with theoretical predictions. The insights obtained here regarding heat dissipation should hold for any mesoscopic system where charge transport is predominantly elastic. Such systems include semiconductor nanowires, two-dimensional electron gases, semiconductor heterostructures, carbon nanotubes, and graphene, among others.

**Methods Summary**

Single-molecule and atomic junctions were created by displacing the NTISTP towards a Au substrate at 5 nm/s and withdrawing from the substrate at 0.1 nm/s after contact formation (indicated by an electrical conductance greater than $5G_0$). The Au substrate was coated with the desired molecules for molecular experiments and was pristine for the atomic junction studies. To obtain the conductance traces a voltage bias of 100 mV is applied and the current is monitored during the withdrawal process. The obtained traces were analyzed by creating histograms to identify the most probable conductance of AMJs. Stable single-molecule junctions with a desired conductance were created by stopping the withdrawal when a conductance plateau with a conductance within 10% of the most probable conductance was obtained. All the experiments were performed in an Ultra-High Vacuum Scanning Probe Microscope at ambient temperature. Further, high-resolution temperature measurements were enabled by a modulation scheme where a time-dependent voltage, $V_M(t)$, consisting of a periodic series of three level voltage pulses $+V_M$, 0 V, $-V_M$ (Fig. S1, supplementary information) was applied to the AMJs while monitoring the thermoelectric voltage output of the NTISTP. The zero-bias transmission functions, shown in Figs. 2-4, were computed with the *ab initio* method described in Ref. 24.

**References and Notes:**

1   Cuevas, J. C. & Scheer, E. *Molecular Electronics: An Introduction to Theory and Experiment*. (World Scientific, 2010).
2   Scheer, E. *et al.* The signature of chemical valence in the electrical conduction through a single-atom contact. *Nature* **394**, 154-157 (1998).
3   Song, H. *et al.* Observation of molecular orbital gating. *Nature* **462**, 1039-1043 (2009).
4   Venkataraman, L., Klare, J. E., Nuckolls, C., Hybertsen, M. S. & Steigerwald, M. L. Dependence of single-molecule junction conductance on molecular conformation. *Nature* **442**, 904-907 (2006).
5   Xu, B. Q. & Tao, N. J. Measurement of single-molecule resistance by repeated formation of molecular junctions. *Science* **301**, 1221-1223 (2003).
6   Ludoph, B. & van Ruitenbeek, J. M. Thermopower of atomic-size metallic contacts. *Phys. Rev. B* **59**, 12290-12293 (1999).
7   Reddy, P., Jang, S. Y., Segalman, R. A. & Majumdar, A. Thermoelectricity in molecular junctions. *Science* **315**, 1568-1571 (2007).
8   Widawsky, J. R., Darancet, P., Neaton, J. B. & Venkataraman, L. Simultaneous determination of conductance and thermopower of single molecule junctions. *Nano Lett.* **12**, 354-358 (2012).
9   Galperin, M., Saito, K., Balatsky, A. V. & Nitzan, A. Cooling mechanisms in molecular conduction junctions. *Phys. Rev. B* **80**, 115427 (2009).
10  Dubi, Y. & Di Ventra, M. Colloquium: Heat flow and thermoelectricity in atomic and molecular junctions. *Rev. Mod. Phys.* **83**, 131 (2011).
11  Karlström, O., Linke, H., Karlström, G. & Wacker, A. Increasing thermoelectric performance using coherent transport. *Phys. Rev. B* **84**, 113415 (2011).
12  Lepri, S., Livi, R. & Politi, A. Thermal conduction in classical low-dimensional lattices. *Phys. Rep.* **377**, 1-80 (2003).

13  Li, N. B. *et al.* Colloquium: Phononics: Manipulating heat flow with electronic analogs and beyond. *Rev. Mod. Phys.* **84**, 1045-1066 (2012).
14  Agraït, N., Untiedt, C., Rubio-Bollinger, G. & Vieira, S. Onset of energy dissipation in ballistic atomic wires. *Phys. Rev. Lett.* **88**, 216803-216806 (2002).
15  Kim, Y., Pietsch, T., Erbe, A., Belzig, W. & Scheer, E. Benzenedithiol: A broad-range single-channel molecular conductor. *Nano Lett.* **11**, 3734-3738 (2011).
16  Huang, Z. F. *et al.* Local ionic and electron heating in single-molecule junctions. *Nature Nanotech.* **2**, 698-703 (2007).
17  Ward, D. R., Corley, D. A., Tour, J. M. & Natelson, D. Vibrational and electronic heating in nanoscale junctions. *Nature Nanotech.* **6**, 33-38 (2011).
18  Ioffe, Z. *et al.* Detection of heating in current-carrying molecular junctions by Raman scattering. *Nature Nanotech.* **3**, 727-732 (2008).
19  Datta, S. *Electronic Transport in Mesoscopic Systems*. (Cambridge University Press, 1995).
20  Lee, W. & Reddy, P. Creation of stable molecular junctions with a custom-designed scanning tunneling microscope. *Nanotechnology* **22**, 485703 (2011).
21  Kiguchi, M., Miura, S., Hara, K., Sawamura, M. & Murakoshi, K. Conductance of a single molecule anchored by an isocyanide substituent to gold electrodes. *Appl. Phys. Lett.* **89**, 213104 (2006).
22  Sivan, U. & Imry, Y. Multichannel landauer formula for thermoelectric transport with application to thermopower near the mobility edge. *Phys. Rev. B* **33**, 551-558 (1986).
23  Paulsson, M. & Datta, S. Thermoelectric effect in molecular electronics. *Phys. Rev. B* **67**, 241403 (2003).
24  Pauly, F. *et al.* Cluster-based density-functional approach to quantum transport through molecular and atomic contacts. *New. J. Phys.* **10**, 125019 (2008).
25  Xue, Y. Q. & Ratner, M. A. End group effect on electrical transport through individual molecules: A microscopic study. *Phys. Rev. B* **69**, 085403 (2004).
26  Malen, J. A. *et al.* Identifying the length dependence of orbital alignment and contact coupling in molecular heterojunctions. *Nano Lett.* **9**, 1164-1169 (2009).
27  Venkataraman, L. *et al.* Single-molecule circuits with well-defined molecular conductance. *Nano Lett.* **6**, 458-462 (2006).
28  Brandbyge, M. *et al.* Quantized conductance in atom-sized wires between two metals. *Phys. Rev. B* **52**, 8499-8514 (1995).
29  Nielsen, S. K. *et al.* Current-voltage curves of atomic-sized transition metal contacts: An explanation of why Au is ohmic and Pt is not. *Phys. Rev. Lett.* **89**, 066804 (2002).
30  Tsutsui, M., Kawai, T. & Taniguchi, M. Unsymmetrical hot electron heating in quasi-ballistic nanocontacts. *Sci. Rep.* **2**, 217 (2012).

**Acknowledgements**

P.R. acknowledges support from the US Department of Energy, Office of Basic Energy Sciences, Division of Materials Sciences and Engineering under award no. DE-SC0004871 (nanofabrication of novel scanning probes), from the National Science Foundation under award no. CBET 0844902 (instrumentation for real-time control) and from the Center for Solar and Thermal Energy conversion, an Energy Frontier Research Center funded by the US Department of Energy, Office of Science, Basic Energy Sciences under award no. DE-SC0000957 (development of a scanning probe microscope). L.A.Z. acknowledges financial support from the Spanish MICINN through grant no. FIS2010-21883. F.P. acknowledges funding through the Carl Zeiss Stiftung, the DFG SFB 767, and the Baden-Württemberg Stiftung. P.R. thanks E. Meyhofer for discussions and comments. P.R. and J.C.C. thank A. Nitzan for discussions. J.C.C. is grateful for the hospitality provided by the Institute for Advanced Studies of the Hebrew University of Jerusalem, where part of this work was carried out.

**Author Contributions**

The experiments were conceived by P.R. and J.C.C. The experiments were performed by W.L. and K.K. The custom-fabricated probed were designed, fabricated and characterized by K.K. and W.J. *Ab initio* charge transport calculations were performed by L.A.Z. and F.P. The manuscript was written by P.R. and J.C.C. with comments and inputs from all authors.

**Author Information** Reprints and permissions information is available at www.nature.com/reprints. The authors declare no competing financial interests. Correspondence and requests for materials should be addressed to P.R. (pramodr@umich.edu) and J.C.C. (juancarlos.cuevas@uam.es).

**Figure 1 | Nanoscale thermocouple probes and atomic and molecular junctions studied in this work.** (**a**) Scanning electron microscope (SEM) image of a NTISTP. The electrodes are false colored. The inset shows a magnified SEM image of the tip. (**b**) Schematic of a junction created between the NTISTP (cross-sectional view) and a Au substrate along with a thermal resistance network that represents the dominant resistances to heat flow. (**c**) Schematics of molecular and atomic junctions along with the structures of the molecules studied. (All schematics not drawn to scale and proportion).

**Figure 2 | Relationship between heat dissipation asymmetries and electronic transmission characteristics in Au-BDNC-Au junctions.** (**a**) Horizontally offset conductance traces (inset) of BDNC junctions along with a histogram obtained from 500 traces. The red line represents a Gaussian fit to the histogram. (**b**) The measured time-averaged temperature rise of the thermocouple ($\Delta T_{TC,\,Avg}$) and the power dissipation in the probe ($Q_{P,\,Avg}$) are shown as a function of the time-averaged total power dissipation in the junction ($Q_{Total,\,Avg}$) for positive and negative biases. The uncertainty at the highest power is 0.6 mK and is less than 0.5 mK for smaller powers. The computationally predicted relationship between $Q_P$ and $Q_{Total}$ is shown by solid lines. The inset shows the measured temperature rise as a function of the magnitude of the applied voltage bias. (**c**) The I-V characteristics of Au-BDNC-Au junctions obtained by averaging 100 individual I-V curves (solid-curve). The shaded region represents the standard deviation of the I-V curves. (**d**) The computed zero-bias transmission function corresponding to the Au-BDNC-Au junction shown in the inset.

**Figure 3 | Heat dissipation asymmetry for Au-BDA-Au junctions. (a - d)** Same as Fig. 2 but for Au-BDA-Au junctions. **(b)** In contrast to Au-BDNC-Au junctions the heat dissipated in the probe is found to be larger for positive biases than for negative ones. The uncertainty on each data point is less than 0.4 mK.

**Figure 4 | No detectable heating asymmetry in Au-Au atomic junctions. (a)** The measured $\Delta T_{TC, Avg}$ and $Q_{P, Avg}$ in Au-Au atomic junctions for positive and negative biases as a function of $Q_{Total, Avg}$. The inset plot shows similar measurements for a larger range of powers. The measured temperature rise is found to be linearly dependent on $Q_{Total, Avg}$ and is independent of the bias polarity within experimental uncertainty (<0.1 mK). **(b)** The computed transmission function corresponding to the Au-Au atomic junction shown in inset-i features a weak energy dependence around the Fermi energy ($E_F$). Inset-ii shows the experimentally obtained I-V characteristics of Au-Au atomic junction created by averaging over 100 independent I-V curves.

## Methods

**Creation of Atomic and Molecular Junctions.** All the AMJs were created between NTISTP and a Au coated substrate by displacing the NTISTP towards a Au substrate (which was coated with the desired molecules in molecular experiments and was pristine in atomic junction experiments) at 5 nm/s and withdrawing from the substrate at 0.1 nm/s after contact formation as indicated by an electrical conductance greater than $5G_0$. To create the desired monolayers 1 mM solutions of BDNC and BDA molecules, obtained commercially from Sigma Aldrich with a purity of ~99%, were created in toluene/ethanol. Subsequently, a Au coated mica substrate (ebeam evaporation) was placed in one of the solutions to self-assemble molecules on the Au surface. After exposing the substrates for 12 hours in a glove box filled with nitrogen gas, they were rinsed in ethanol and dried in nitrogen gas. For the experiments involving Au-Au atomic junctions the Au coated substrates were cleaned in UV-Ozone to eliminate any organic contamination on the surface. The NTISTPs were also UV-Ozone cleaned in all studies and loaded into the UHV scanning probe microscope instrument. The measurement of electrical current was performed using a current amplifier (Keithley 428), whereas thermoelectric voltage measurements were performed using a voltage amplifier (Stanford Research System 560). All the data was collected at a sampling frequency of 2 kHz using a data acquisition system (National Instruments 6281). The approach, withdraw, and hold sequences were accomplished by employing a real-time controller (National Instruments PXI8110).

**Measurement of $\Delta T_{TC,\ Avg}$ Using a Modulation Scheme.** High-resolution temperature measurements are enabled by a modulation scheme where a time-dependent voltage, $V_M(t)$, consisting of a periodic series of three level voltage pulses $+V_M$, 0 V, $-V_M$, (Fig. S1 of the SI) is applied. In all the experiments performed in this work, the period ($T_P$) of the voltage pulses was chosen to be ~0.08 seconds (1/12.25 Hz). The selected modulation frequency is found to optimize the signal to noise ratio and is experimentally feasible due to the small thermal time constants (~10 μs) of the micrometer sized NTISTPs, which enables high fidelity tracking of temperature changes. The applied $V_M(t)$ results in both a modulated current ($I_M(t)$ (see Fig. S1) and a modulated temperature change of the thermocouple ($\Delta T_{M, TC}(t)$) due to Joule heating. Using the equation at the bottom of Fig. S1, the temperature rise corresponding to a positive bias $\Delta T_{TC, Avg}(+V_M)$ or a negative bias $\Delta T_{TC, Avg}(-V_M)$ can be directly related to the modulated thermoelectric voltage output ($\Delta V_{M, TC}(t)$) of the thermocouple. In probing heat dissipation in AMJs we applied the modulated voltage signal with an appropriately chosen amplitude $V_M$ for a period of ~5 seconds to each AMJ. The resulting thermoelectric voltage signal $\Delta V_{M, TC}(t)$ was simultaneously recorded. This was repeated on ~100 junctions to collect data for ~500 seconds for each $V_M$. The obtained data was concatenated and analyzed to estimate $\Delta T_{TC, Avg}$ corresponding to positive and negative biases as described above. This modulation scheme enables temperature measurements with sub milli-Kelvin resolution as described in the SI. The total time-averaged power dissipation ($Q_{Total, Avg}$), at each bias, was obtained by using the 500 second long data corresponding to each $V_M$. Specifically, the data (measured current and known applied bias) was used to first compute the total power dissipated ($Q_{Total}(+V_M/-V_M)$) at positive and negative biases. Subsequently, $Q_{Total, Avg}(+V_M/-V_M)$ was obtained by dividing $Q_{Total}(+V_M/-V_M)$ by the total time during which a positive bias ($+V_M$) or negative bias ($-V_M$) was applied (~500/3 seconds). The amplitudes ($V_M$) of the three level voltage pulses employed in our studies were chosen to be 30 mV, 43 mV, 52 mV, 60 mV, and 67 mV for Au-Au junctions, 0.74 V, 0.95 V, 1.08 V, 1.18V, and 1.27 V for Au-BDNC-Au junctions, and 0.44 V, 0.58 V, 0.68

V, 0.76 V, and 0.82 V for Au-BDA-Au junctions. Representative traces obtained in the experiments are shown in section 6.3 of the SI.

**Estimating $Q_{P,\ Avg}$ from the Measured $\Delta T_{TC,\ Avg}$.** To relate the temperature rise of the thermocouple to the time-averaged power dissipation in the probe $Q_{P,\ Avg}$ it is necessary to quantify the thermal resistance of the NTISTP. To elaborate, consider the resistance network shown in Fig. 1b, where the thermal resistances to heat flow in the probe ($R_P$), junction ($R_J$), and the substrate ($R_S$) are identified. $R_P$ was experimentally determined to be 72800 ± 500 K/W (see SI). The thermal resistances of AMJs ($R_J$) are estimated to be at least $10^7$ K/W for all the AMJs studied here (see SI for more details). Thus, $R_J \gg R_P$ and therefore, $\Delta T_{TC,\ Avg}$ depends only on the power dissipated in the tip and is unaffected by the heating in the substrate. Thus, from a knowledge of $\Delta T_{TC,\ Avg}$ and $R_P$, the time-averaged power dissipation, $Q_{P,\ Avg}$, can be estimated as $Q_{P,\ Avg} = \Delta T_{TC,\ Avg} / R_P$.

**Computation of the Transmission Function.** The zero-bias transmission functions shown in the manuscript were computed with the *ab initio* method described in detail by us before[24]. It is based on a combination of non-equilibrium Green's function techniques and density functional theory (DFT) and was implemented in the quantum-chemistry software package Turbomole. More details can be found in the SI.

**Computing the Relationship between $Q_P$ and $Q_{Total}$.** We computed the power dissipated in the probe $Q_P(V)$ and the total power dissipated in the junction $Q_{Total}$ ($Q_P(V) + Q_S(V) = Q_{Total}(V)$) using equation (1) and the zero-bias transmission curves of the molecular junctions (shown in Figs. 2d and 3d). Subsequently, $Q_P$ was plotted as a function of $Q_{Total}$ as the relationship between $Q_P$ and $Q_{Total}$ is robustly predicted by our calculations (see SI for details).

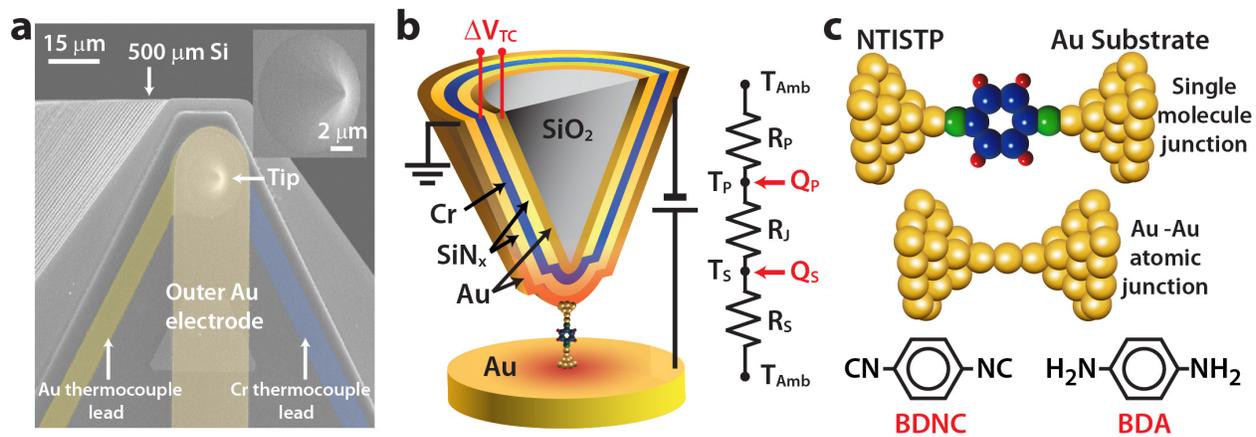

**Figure 1**

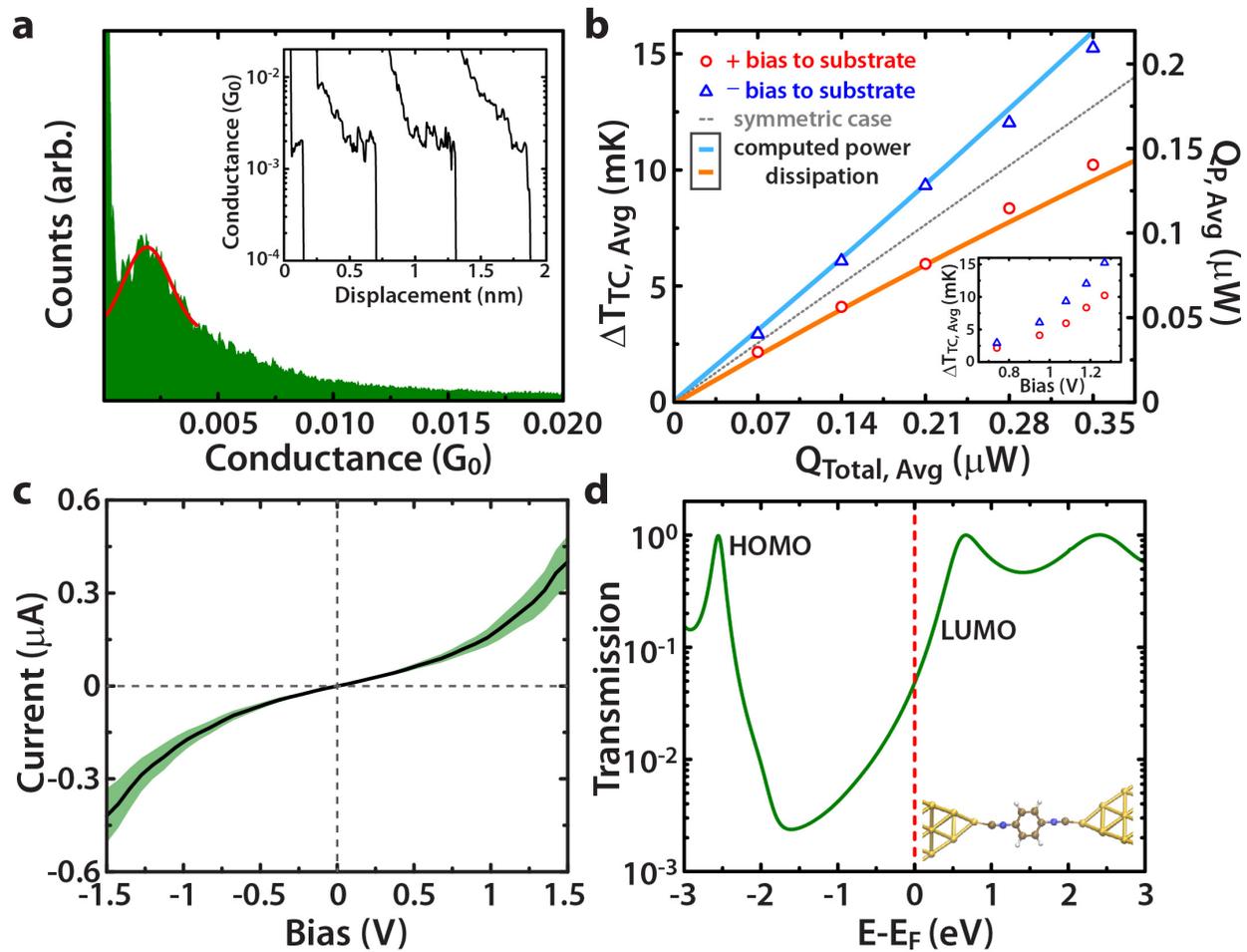

Figure 2

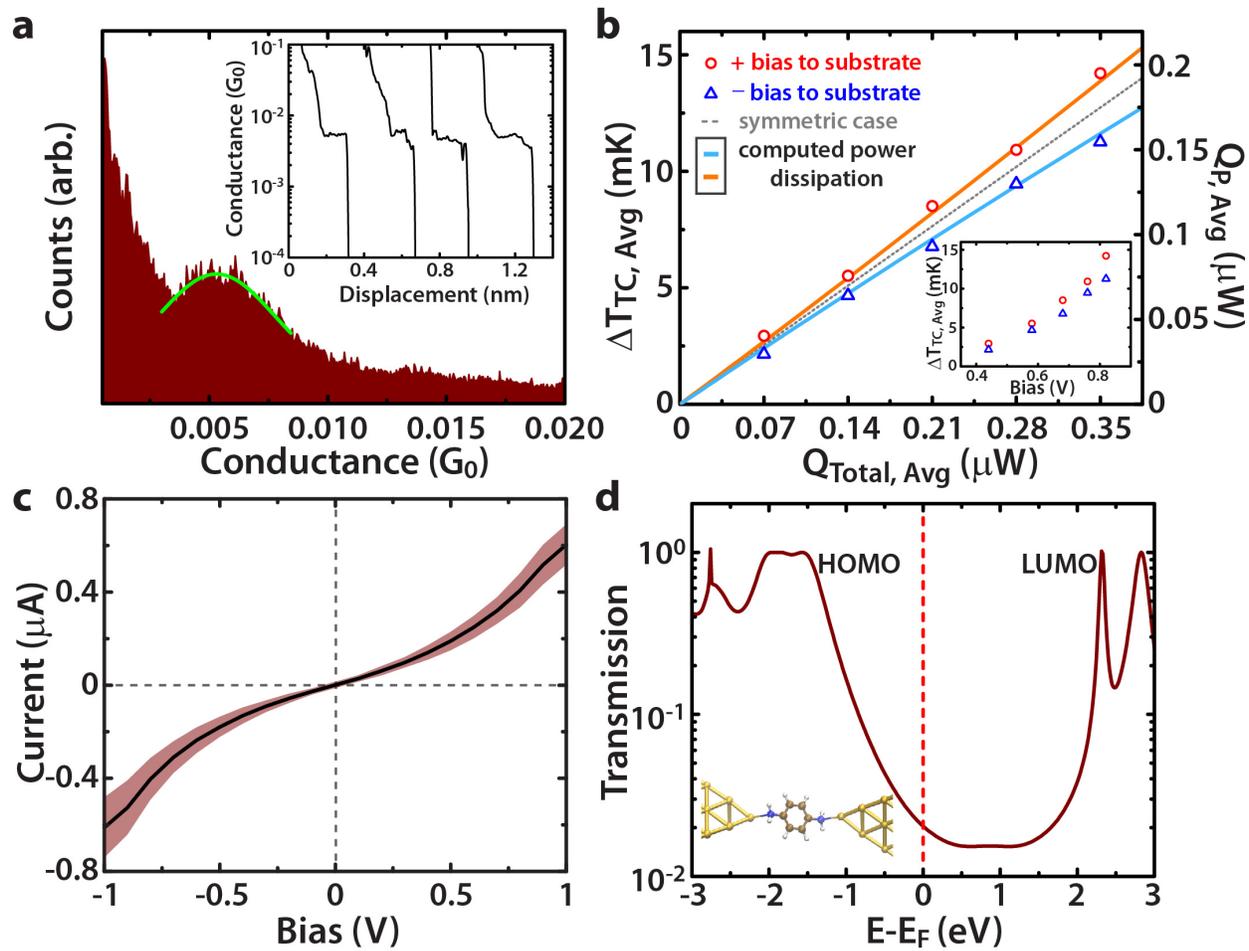

**Figure 3**

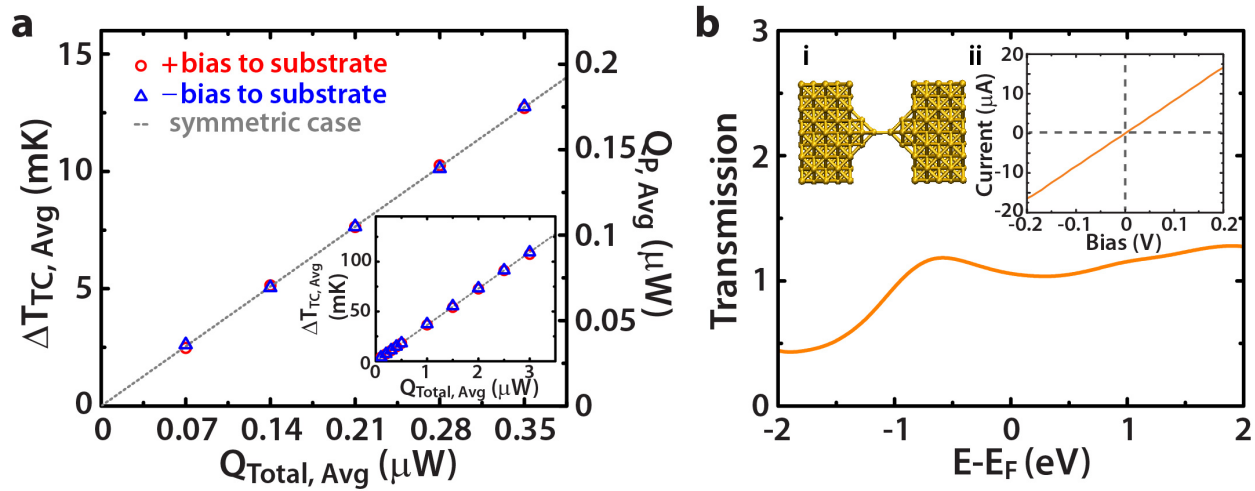

**Figure 4**